\documentclass[%
reprint,amsmath,amssymb,aps,
]{revtex4-2}

\usepackage{graphicx}
\usepackage{dcolumn}
\usepackage{bm}
\usepackage{float}
\usepackage{amsmath}
\usepackage[toc,page]{appendix} 
\usepackage[thinc]{esdiff}
\usepackage{blindtext}
\usepackage{multirow,booktabs}
\usepackage{dcolumn}
\usepackage{bm}
\usepackage{units}
\usepackage[version=4]{mhchem}
\usepackage{booktabs,chemformula}
\usepackage{color}
\usepackage[bottom]{footmisc}
\usepackage{xcolor}
\usepackage{media9}

\begin{document}
\raggedbottom

\title{A Morphologically Self-Consistent Phase Field Model for the Computational Study of Memristive Thin Film \textit{Current-Voltage} Hysteresis}

\author{John F. Sevic}
\affiliation{Department of Electrical, Computer, and Software Engineering, Embry-Riddle Aeronautical University, Prescott, AZ, USA}
\author{Ambroise Juston}
\affiliation{Department of Aeronautical Engineering\linebreak Embry-Riddle Aeronautical University, Prescott, AZ, USA}
\author{Nobuhiko P. Kobayashi}
\affiliation{Department of Electrical and Computer Engineering\linebreak  University of California, Santa Cruz, CA, USA}

\begin{abstract}
A multiphysics phase field model is used for the computational study of memristive thin film morphology and \textit{current-voltage} hysteresis. In contrast to previous computational methods, no requirements are made on conducting filament geometry. Our method correctly predicts conducting filaments evolve on thermodynamic paths that are energetically favored due to stochastic structural and chemical variations naturally occurring at the atomic-level, due to both latent and intentional fabrication effects. These results have significant implications for the computational design of a broad class of memristive thin films, enabling practical wafer-scale mapping, uniformity, and endurance analysis and optimization.
\end{abstract}


\maketitle

\section{\label{sec:level1}Introduction}

Memristive thin films owe to their fundamental non-volatile memory operation anomalous charge transport, by which we mean thin films nominally classified by band theory as electrically insulating. In these thin films, memristive electric conduction manifests as \textit{current-voltage} hysteresis, making them ideal candidates for next-generation solid-state memory due to reversible bistable persistence. Memristive thin films have an interesting and rich history, initially misunderstood and later, as fabrication techniques matured and analytical understanding improved, treated to deliberate preparation methods to specifically and systematically exploit production of hysteresis \citep{Mott1940}\cite{Hickmott62}\citep{Mott1968}\citep{Waser2007}\citep{Strukov2008}.

Memristive electric conduction is produced by many physical behaviors, and a variety of models have been proposed for the computational study of these memristive thin films to optimize their electrical behavior and performance. Broadly speaking, models are classified based on physical abstraction. The simplest models adopt an equivalent electrical network, followed by adoption of continuum transport, implementation of particle motion tracking, and formulation from first-principles. Models based on variational methods have been also been proposed. Each of these methods can be further classified based on complexity and computational efficiency, often with a trade-off between them. The simplest memristor models are often applied to the study of large-scale neuromorphic networks \citep{Gibson2016}. Continuum models, usually based on a drift-diffusion transport formulation, benefit from a well-understood and mature computational framework and replicate observed memristive \textit{current-voltage} behavior as well as providing a reasonable picture of ionic motion and conducting filament formation \citep{Nardi2012}\citep{Larentis2012}\citep{Kim2013}\citep{Sevic2018}. The continuum model, however, depends vitally on \textit{a priori} definition of an idealized axisymmetric conducting filament geometry that is inconsistent with experimentally observed morphology. Both molecular dynamics and first-principle approaches have been proposed, often in tandem, but these are computationally intensive and not suitable for wafer-scale analysis and optimization \citep{Shen2018}\citep{Fyrigos2021}\citep{Funck2021}.

A computationally efficient model that does not require \textit{a priori} geometry assumptions, and instead yields spontaneous conducting filament production self-consistently with the inherently stochastic influence of intentional and random defects of as-fabricated memristive thin films, offers significant advantages. A phase field model is one such approach \citep{Sevic2019}\citep{Sevic2023}. In this paper, we extend our multiphysics phase field model to demonstrate \textit{current-voltage} hysteresis of transition metal oxide thin films and visualization of spontaneous conducting filament formation with morphology consistent with experimental observations. Our model provides a scalable and efficient method of wafer-scale mapping, uniformity, and endurance analysis and optimization.

\section{\label{sec:level2}Memristive Electric Conduction}

Memristive electric conduction exhibited by appropriately prepared thin films is produced by a broad variety of physical behaviors, including phase change, quantum tunneling, field assisted transport, and various valance-related mechanisms \citep{Pan2010}\citep{Miao2011}\citep{Strachan2010}\citep{Xu2008}. In many instances, memristive electric conduction is substantially influenced by simultaneous combination of these behaviors, as in various forms of hopping \footnote{By hopping we mean both discrete jumps in energy and space.}\citep{Wilson32}\citep{Mott1938}\citep{Baranovskii14}. With phase change thin films, electron-pair interaction or crystallographic configuration modulation yields reversible bistability between an amorphous, insulating state, and a crystalline, metallic state, by electrothermal stimulus \citep{Zhang2021}\citep{Juan2016}.  Quantum tunneling contributes to both hopping within bulk the thin film and at the metal-dielectric interface formed by the electrodes. By field assisted transport we mean specifically charge transport due to presence of states for thermally excited mobile charges simultaneously accelerated by an external electric field, such as Frenkel or Schottky bulk and interface defects, respectively.

Memristive electric conduction produced by transition metal oxides is due to various valance-related mechanisms, such as redox-driven reactions based on electrochemical formation or dissolution of various ionic species, such as silver or copper. These thin films produce charge transport from both motion and redistribution of oxygen vacancies and concurrent redox reactions and depends on interface phenomena with the metallic electrodes on the thin film \citep{Yang2009}\citep{Kim2015}\citep{Juan2017}\citep{Kim2022a}\citep{Disha2024}.Associated with valance change mechanisms is spontaneous production of a characteristic nanoscale morphology, often referred to as a conducting filament, reflecting ionic motion and redistribution within the host thin film \citep{Yang2012}\citep{Ahmed2018}\citep{Bejtka2020}. The \textit{current-voltage} behavior of these charge transport phenomena is controlled by specific thin film fabrication techniques, producing both intentional and random lattice imperfections that establish a local density of states (LDOS) that yields random distributions of states in energy and space, presenting a challenge to analytic and computational treatment. These intentional and random lattice imperfections nevertheless have a direct impact on the \textit{current-voltage} hysteresis we seek to understand and optimize for memristor operation.

Necessary for the study of valance-related electric conduction, then, is a model that maps the effect of these intentional and random lattice imperfections to energy and spatial states within the thin film. Prediction of these states can be implemented from first-principle density functional theory, though it is computationally intensive to include random effects \citep{Fyrigos2021}\citep{Funck2021}. Here we propose instead a computationally efficient and scalable model derived from a thermodynamic variational principle that is inherently compatible with randomness associated as as-fabricated memristive thin films.

Charge density is defined by LDOS and a charge carrier energy distribution function in the usual manner as

\begin{equation}
\label{DOS_integral}
n(\vec{r}) \;= \int_{\Delta E} dE \times G(\vec{r},E,t_{0}) \times f(\vec{r},E)
\end{equation}

where the $\Delta E$ is the energy range of interest. The LDOS at location $\vec{r}$ is defined as a random process by $G(\vec{r},E,t_{0})$, and the variable $t_o$ indicates these states are established at this time as an initial condition. For memristive thin films is it reasonable to to assume a Boltzmann distribution, since charges are largely non-interacting. In this case, the probability of a charge carrier at location $\vec{r}$ within the thin film possessing energy $E(\vec{r})$ is approximated as $f(\vec{r},E) \approx exp[-\beta E(\vec{r}) ]$ where $\beta$ is inverse thermal energy.

To compute steady-state charge density, we exploit the inherently stochastic LDOS of as-fabricated memristive thin films by treating the problem statistically. That is, we ask what ensemble of states established by $G(\vec{r},E,t_{0})$ under the influence of an external electric field self-consistently minimizes the Gibbs free energy of the entire system subject to spontaneous conducting filament production. To identify this minimum Gibbs free energy, we employ a variational formulation based on Landau mean field theory in which a unitless parameter, $0 \leq \phi(\vec{r}) \leq 1$, represents the combined influence of lattice imperfections on spatial and energy levels and charge carrier energy distribution \citep{Landau1980}. In our approach, we assume that states are sufficiently delocalized that we can ignore long range correlation and that our system can be treated as a statistical ensemble based on a random LDOS reflecting the influence of intentional and random lattice defects, a reasonable approximation for amorphous thin films.

The minimum energy ensemble is obtained by identification of a $\phi(\vec{r})$ that minimizes the Gibbs free energy $F[\phi(\vec{r})]$ expressed as

\begin{equation}
\label{Gibbs}
F\Bigl[\phi(\vec{r})\Bigr] = U\Bigl[\phi(\vec{r})\Bigr] - T \times S\Bigl[\phi(\vec{r})\Bigr]
\end{equation}

where $U[\phi(\vec{r})]$ and $S[\phi(\vec{r})] \approx kT\log(g_{L})$ are the internal energy and entropy of the system in the state identified by $\phi(\vec{r})$, the order parameter  \citep{Provatas2010}. The Gibbs free energy in this case is the total energy of the system, including bulk energy of the thin film,  electrostatic energy, and now, importantly, interface energy since our approach stipulates spontaneous conducting filament production. We can form a discrete approximation to charge density Equation \ref{DOS_integral} in the $\phi(\vec{r})$-domain as

\begin{equation}
\label{DOS_sum}
n(\vec{r},\phi) \;\approx\; \sum_{i} G(\vec{r},\phi,t_{0}) \times exp \Bigl[\eta(\phi-\phi_o) \Bigr]
\end{equation}

where $n(\vec{r},\phi)$ is the approximate charge density at location $\vec{r}$ in the thin film and \textit{i} is an index over the minimum Gibbs energy ensemble produced by $\phi(\vec{r})$. Here $E(\vec{r})$ and $\beta$ have been replaced by order parameter $\phi(\vec{r})$ and empirical constants $\eta$ and $\phi_o$ Since choosing the range of order parameter $\phi(\vec{r})$ and attaching physical meaning to its extremum is arbitrary, $\eta$ maps the likelihood of a charge carrier of specific energy to the difference of order parameter $\phi(\vec{r})$ and $\phi_o$, where $0 \leq \phi_o(\vec{r}) \leq 1$. A similar process was carried out on $G(\vec{r},E,t_{0})$ for mapping to the $\phi(\vec{r})$-domain. With our model, $\phi(\vec{r}) \approx 1$ thus represents a relatively high electrical conductivity and $\phi(\vec{r}) \approx 0$ represents a relatively low electrical conductivity. 

\section{\label{sec:level2}Phase Field Model}

\begin{figure}[h]
\centering
\includegraphics[scale=0.23, angle=0]{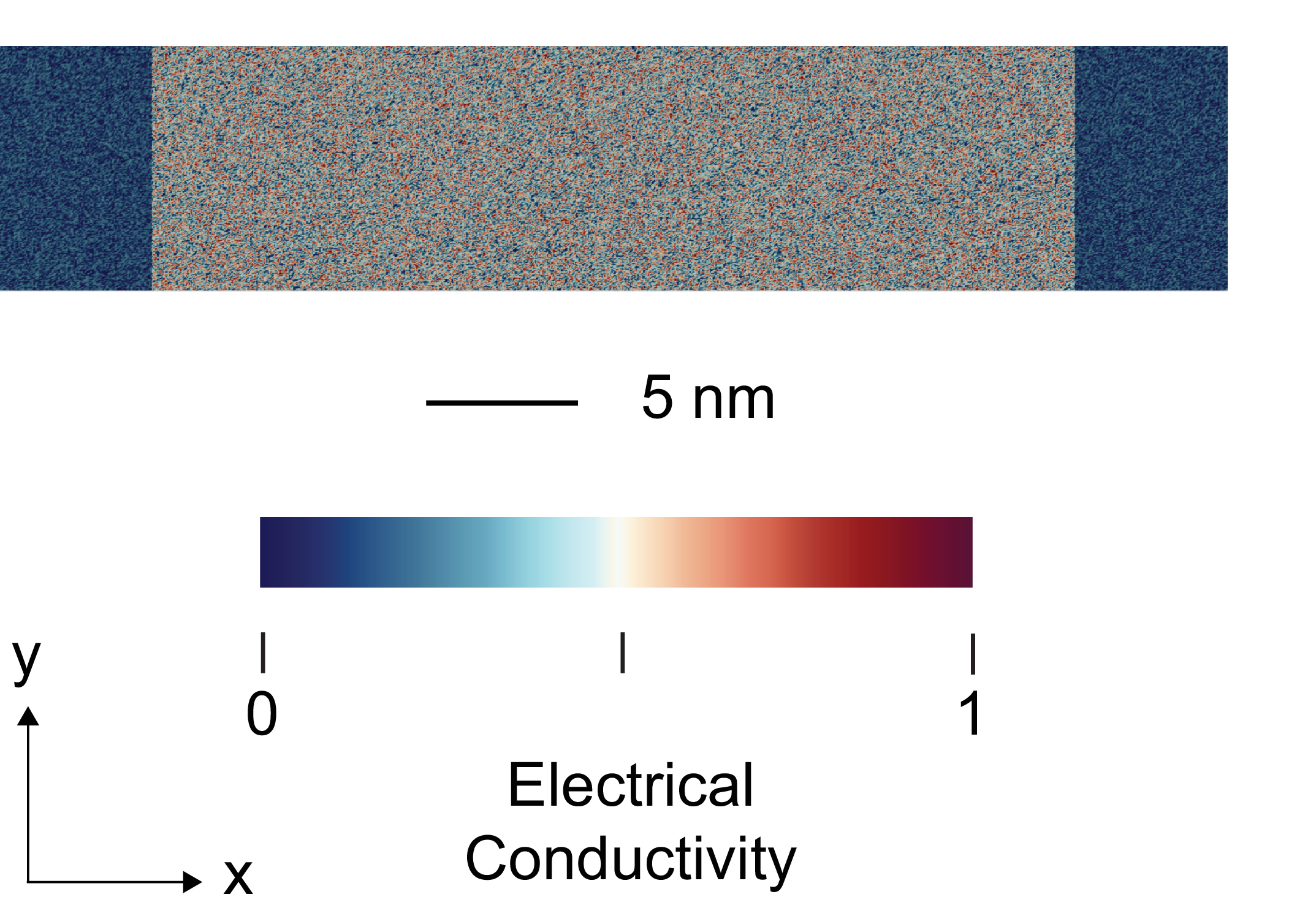}
\caption{Cross section and initial normalized electrical conductivity of our thin film phase field model. The active region is defined from  $x_1 = 5\,$nm and $x_1 = 35\,$nm. The top and bottom electrodes defined by the active region are assumed to be ideal electrical contacts. The regions on either side of the active region are electrically insulating buffer zones.}
\label{deviceStructure}
\end{figure}

Steady-state conducting filament morphology and \textit{current-voltage} behavior is identified by solving for the order parameter $\phi(\vec{r})$ that minimizes the Gibbs free energy Equation \ref{Gibbs} subject to appropriate initial conditions. Figure \ref{deviceStructure} shows the initial normalized electrical conductivity map of our memristive thin film model established by $G(\vec{r},\phi,t_{0})$. Random variation in electrical conductivity represents both latent and intentional artifacts of fabrication. The regions on either side of the active region are electrically insulating buffer zones. The top and bottom electrodes are assumed to be ideal electrical contacts.

\begin{figure}[h]
\centering
\includegraphics[scale=0.4, angle=0]{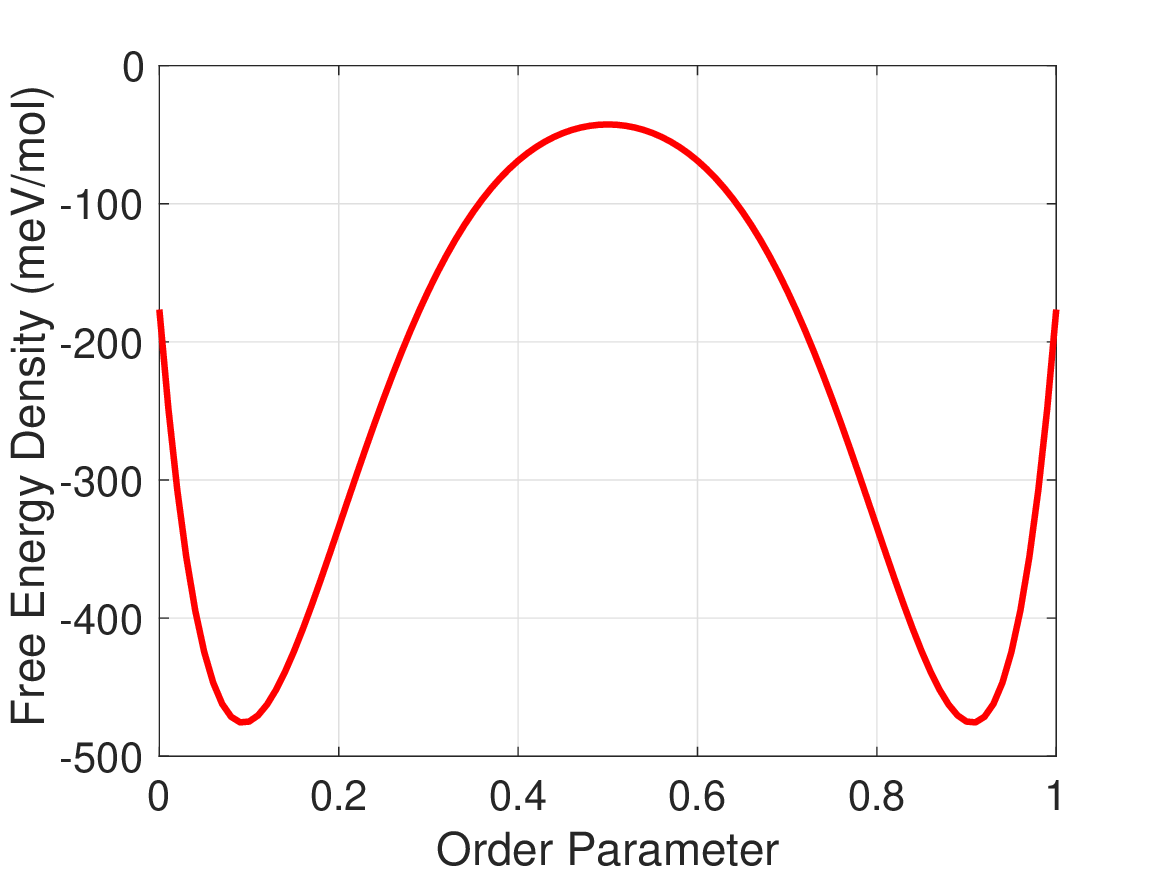}
\caption{Free energy density as a function order parameter $\phi(\vec{r})$. This free energy density function represents energy states corresponding to tantalum oxide-based memristive thin films from our earlier work \citep{Juan2017}. The two equilibrium states represent the propensity of charges to form equilibrium clusters and assumes the energy of formation and dissolution are identical. Well depth is an approximation for the energy required to change from each charge density equilibrium state.}
\label{FEDF}
\end{figure}

To model electrical conductivity reversible bistability due to oxygen vacancy motion and redistribution within the memristive thin film of Figure \ref{deviceStructure}, we employ a double-well free energy density function \cite{Sevic2019}\cite{Sevic2023}. This free energy density function approximates energy states corresponding to tantalum oxide-based memristive thin films from our earlier work, and is illustrated by Figure \ref{FEDF} \citep{Juan2017}. For bistable charge density modulation, we anticipate the free energy density to be a function of order parameter $\phi(\vec{r})$ and in our model this is expressed analytically by Equation \ref{DoubleWellPotential} as

\begin{equation}
\label{DoubleWellPotential}
f_{bulk}(\phi) = a_{0} + a_{2}(\phi) + a_{4}(\phi) + a_{6}(\phi)
\end{equation}

where $a_{2}(\phi)$, $a_{4}(\phi)$, and $a_{6}(\phi)$ are even functions of $\phi(\vec{r})$ and $f_{bulk}(\phi)$ is in meV/mol \footnote{For our model the following appropriately scaled even-order polynomial functions in order parameter $\phi$ are used. These functions are based on our earlier electrothermal phase field model but held constant at 300 K absolute temperature for the present isothermal treatment of valance change memristive electric conduction. The functions are $a_{2}(\phi) = -0.065\times\large[4\phi - 2 \large]^2$, $a_{4}(\phi) = 0.70\times\large[4\phi - 2 \large]^4$, $a_{6}(\phi) = 0.25\times\large[4\phi - 2 \large]^6$.}. The two equilibrium states represent the propensity of charges to form equilibrium clusters and assumes the energy of formation and dissolution are identical. Well depth is an approximation for the energy required to change from each charge distribution equilibrium state. Because order parameter $\phi(\vec{r})$ is a function of location so to will be the instantaneous free energy density.

To identify order parameter $\phi(\vec{r})$ that minimizes Gibbs free energy Equation \ref{Gibbs}, we construct an energy functional composed of free energy density, electrostatic potential energy density, and interface energy density expressed as

\begin{equation}
\label{FreeEnergyFunctional}
F(\phi) = \int_R d\vec{r} \times \bigg[ f_{bulk}(\phi) + \frac{\kappa}{2} \nabla^2 \phi + g_{elec}(\phi,V)  \bigg]
\end{equation}

with the integration over region $R$, the entire thin film of Figure \ref{deviceStructure}. The electrostatic energy density term is due to the external electric bias applied between the top and bottom electrodes of the thin film of Figure \ref{deviceStructure} and is expressed as $g_{elec}(\phi,V) = \phi(\vec{r})\times V(\vec{r})\times q$ where $V(\vec{r})$ is the electric potential at a point in $(x,y)$-plane of Figure \ref{deviceStructure} and $q$ is electronic charge. Interface energy term $\kappa$ represents the propensity of spontaneous conducting filament formation. The first variation of the free energy functional Equation \ref{FreeEnergyFunctional} yields

\begin{equation}
\label{PhaseFieldPDE}
\frac{\partial \phi(\vec{r})}{\partial t} =  M\nabla^2 \bigg[ \frac{\partial f_{bulk}(\phi)}{\partial \phi} - \kappa\nabla^2 \phi(\vec{r}) - V(\vec{r}) \bigg]
\end{equation}

where $M$ is the mobility of the phase field flux and assumed uniformly constant over $R$. This is a Cahn-Hilliard phase field equation in order parameter $\phi(\vec{r})$ for our model \cite{Cahn1958}. Because we are interested only in the steady-state charge distribution, charge conservation is enforced by the Gauss law

\begin{equation}
\label{ElectronPDE}
\nabla^{2} V(\vec{r}) = \frac{n(\vec{r},\phi)}{\epsilon_r\epsilon_o}
\end{equation}

where $\epsilon_r$ is the static relative electric permittivity of pentavalent tantalum oxide, assumed to be uniform throughout the thin film, and $\epsilon_o$ is electrical permittivity in vacuum.

Model parameters are defined in Table I, and are based on model constants and material properties from our previous work and work similar to ours; the relevant references for each parameter are shown in the right column.  The phase field mobility, $M$, establishes charge carrier relaxation time and for our model this is on the order of 1 $\mu$s. The interface energy density, $\kappa$, is established to enforce conducting filament morphology consistent with experimental data \cite{KKS}. Since our model is calibrated for transition metal oxides of Type II electric conduction, we therefore anticipate a nonlinear \textit{current-voltage} response for the positive-going bias and an approximately ohmic response for the negative-going bias. Based on this physical reasoning, emission parameter and mean order  parameter $\eta$ and $\phi_o$ have been set to 1000 and 0.8, respectively, producing optimum \textit{current-voltage} hysteresis.

\begin{table}[h]
\label{MaterialTable}
\caption {Electrical and phase field parameters and material properties used in our phase field model Equations \ref{PhaseFieldPDE} and \ref{ElectronPDE}. The relevant references for each parameter are shown in the right column.}
\begin{ruledtabular}
\begin{tabular}{ccccc}
\addlinespace[0.5em]
Parameter & Value & Units & Reference\\
\addlinespace[0.5em]
\hline
\addlinespace[0.8em]
$\eta$ & 1000 & Unitless & See text. & \\
\addlinespace[0.8em]
$\phi_{o}$ & 0.8 & Unitless & See text. & \\
\addlinespace[0.8em]
$\kappa$ & 1.0 & $\dfrac{eV}{nm^{2}}$&\citep{Sevic2019}& \\
\addlinespace[0.8em]
$M$ & 100 & $\dfrac{nm}{J \times ns}$&\citep{Sevic2019}& \\
\addlinespace[0.8em]
$\epsilon_r$ & 25 & Unitless &\cite{Sevic2019}& \\
\addlinespace[0.8em]
$\mu_{o}$ & 100 & $\dfrac{nm}{V \times ns}$&\cite{Sevic2023}& \\
\addlinespace[0.8em]
\end{tabular}
\end{ruledtabular}
\end{table}

Order parameter $\phi(\vec{r})$ is computed by self-consistently solving Equations \ref{PhaseFieldPDE} and \ref{ElectronPDE} with the Multiphysics Object-Oriented Simulation Environment (MOOSE) multiphysics solver at a specified bias $V_o$ \cite{gaston2009moose}. These equations are discretized by an adaptive finite element meshing algorithm and solved by a time-domain Newton method to provide steady-state conducting filament morphology and current density \cite{Tonks2010}\cite{petsc}\cite{libMeshPaper}. To compute the \textit{current-voltage} response, at each $V_o$ step, the initial conditions are established by the previous solution. The total current is computed by integrating current density at the top edge of the thin thin film of Figure \ref{deviceStructure} with electrical conductivity defined as $\sigma(\vec{r},\phi) = n(\vec{r},\phi) \times \mu_{o} \times q$, and assumes its usual continuum form of

\begin{equation}
\label{current}
i \approx q \times \mu_{o} \int_{x_1}^{x_2} dx \times n_y \times E_y
\end{equation}

where $x_1 = 5\,$nm and $x_1 = 35\,$nm define an ideal electrode over memristive active region. The charge density and electric field are evaluated at this same edge, denoted as $n_y$ and $E_y$ respectively, and only the \textit{y}-component of vector electric field $\vec{E}(\vec{r})$ is used in computing current. Under the assumption of an ideal electrical interface to the thin film we believe these are reasonable approximations. For an electrical mobility model, we assume the charge enclosed by the conducting filament responds dynamically to vector electric field $\vec{E}(\vec{r}) = -\nabla V(\vec{r})$ with constant mobility $\mu_{o}$ based on our previous phase field model.

\section{\label{sec:level1}Discussion}

The computational \textit{current-voltage} behavior of our memristive electric conduction phase field model is evaluated by applying bias voltage $V_o$ as a positive-going logarithmic sweep of 13 steps from +100 $\mu$V to +100 mV applied between the top edge and bottom edge of the thin film of structure of Figure \ref{deviceStructure}. At each bias voltage, a self-consistent transient solution to Equations \ref{PhaseFieldPDE} and \ref{ElectronPDE} is performed, with steady-state convergence reached when the change in absolute value of the total Gibbs free energy of the thin film is less than 0.1$\%$ between adjacent iterations. Once the +100 mV solution has reached steady-state, a negative-going logarithmic bias voltage sweep of 12 steps from +100 mV to +100 $\mu$V is made under identical convergence criteria. The initial condition for each voltage step is established by the previous voltage step, except at the starting voltage of +100 $\mu$V, which uses initial conditions of an as-fabricated thin film shown in Figure \ref{deviceStructure}. To evaluate the full range of performance of our model, this process is repeated by an identical logarithmic sweep from -100 $\mu$V to -100 mV and back. A single-point simulation was also performed for a bias of 0 V.

\begin{figure}
\centering
\includegraphics[scale=0.45, angle=0]{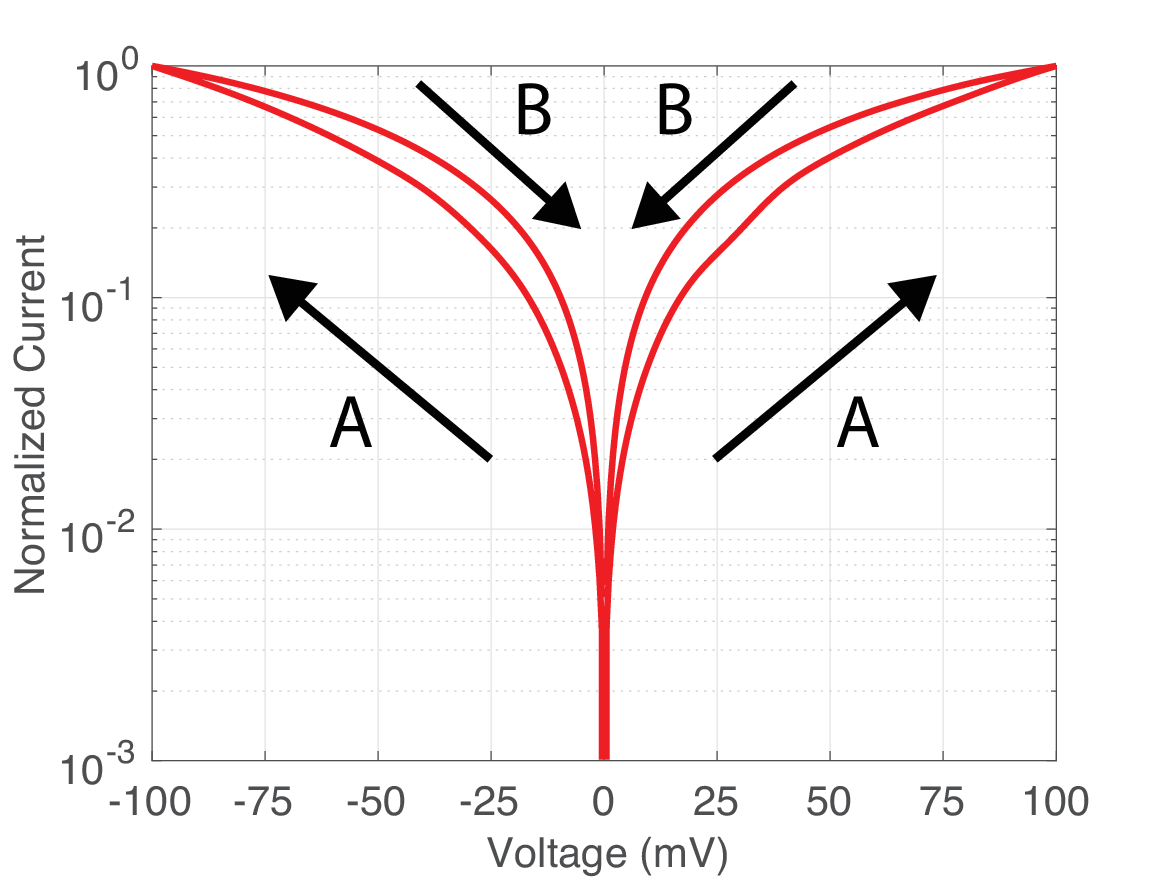}
\caption{The computational \textit{current-voltage} response of our memristive electric conduction phase field model evaluated by applying a positive-going logarithmic bias voltage sweep of 13 steps from +100 $\mu$V to +100 mV applied between the top edge and bottom edge of the thin film of structure of Figure \ref{deviceStructure} and then a negative-going logarithmic voltage ramp of 12 steps from +100 mV to +100 $\mu$V. The process is repeated by application of identical sweep with a negative bias voltage. The positive-going A direction corresponds to the high resistance state and the negative-going B direction corresponds to the low resistance state. Because the initial conditions of positive and negative biases are identical, specifically LDOS, the expected even symmetry \textit{current-voltage} is produced.}
\label{IV}
\end{figure}

Figure \ref{IV} illustrates the computational \textit{current-voltage} response produced by our phase field model corresponding to the initial conditions illustrated by Figure \ref{deviceStructure}. Initially, under positive-going low bias conditions, oxygen vacancy energy is predominantly thermal and is therefore constrained to the lower energy states prescribed by $G(\vec{r},E,t_{0})$. This response is consistent with moderately nonlinear Type II memristive electric conduction. In this condition there is little oxygen vacancy motion and redistribution, but as bias $V_o$ increases toward 100 mV there is substantial charge motion and redistribution due to the potential energy provided by this voltage. Evolution along this positive-going trajectory is denoted by Arrow A on the \textit{current-voltage} response and represents the high resistance state for positive and negative bias.

Because low energy states are occupied first, these states are only available for conduction after an oxygen vacancy has overcome the energy barrier established by the FEDF, representing an activation energy. Therefore, neighboring mobile charges are forced to take an alternative physical path as two charges cannot simultaneously occupy an energy state at the same location. Because of this, clustering occurs, both transverse and longitudinally with respect to the conducting filament axis, as charges aggregate in space and other charge must seek equilibrium elsewhere. This behavior is expected to occur in areas where lower energy states occupy first and where oxygen vacancies are injected at the electrode-dielectric interface. As clusters expand due to full occupancy of states at maximum bias, they eventually connect to form a continuous conducting filament between the top and bottom electrodes of the thin film. This path, denoted by Arrow B in Figure \ref{IV},  reflects evolution to the low resistance state of our memristive thin film, for positive and negative bias, and compares favorably with valance change electric conduction exhibited by transition metal oxide memristive thin films \citep{Yang2009}\citep{Kim2015}\citep{Juan2017}\citep{Kim2022a}\citep{Disha2024}. The maximum high resistance to low resistance ratio produced by our model is $\approx$ 10 and occurs at $\approx \pm$25 mV. 

To evaluate consistency of our method, note that because the initial conditions of positive and negative biases are identical, specifically LDOS, the expected even symmetry of the $current-voltage$ response is produced. Figure \ref{IV} therefore clearly suggests that memristive thin films that are initially structurally symmetric would never exhibit asymmetric \textit{current-voltage} characteristics with respect to zero bias voltage. This is a restatement of evaluation of the flat interface discussed in our earlier work, in which no conducting filaments are produced \footnote{See specifically panels (a) and (b) Figure 4 of \citep{Sevic2019}.}.

\begin{figure}
\centering
\includegraphics[scale=0.23, angle=0]{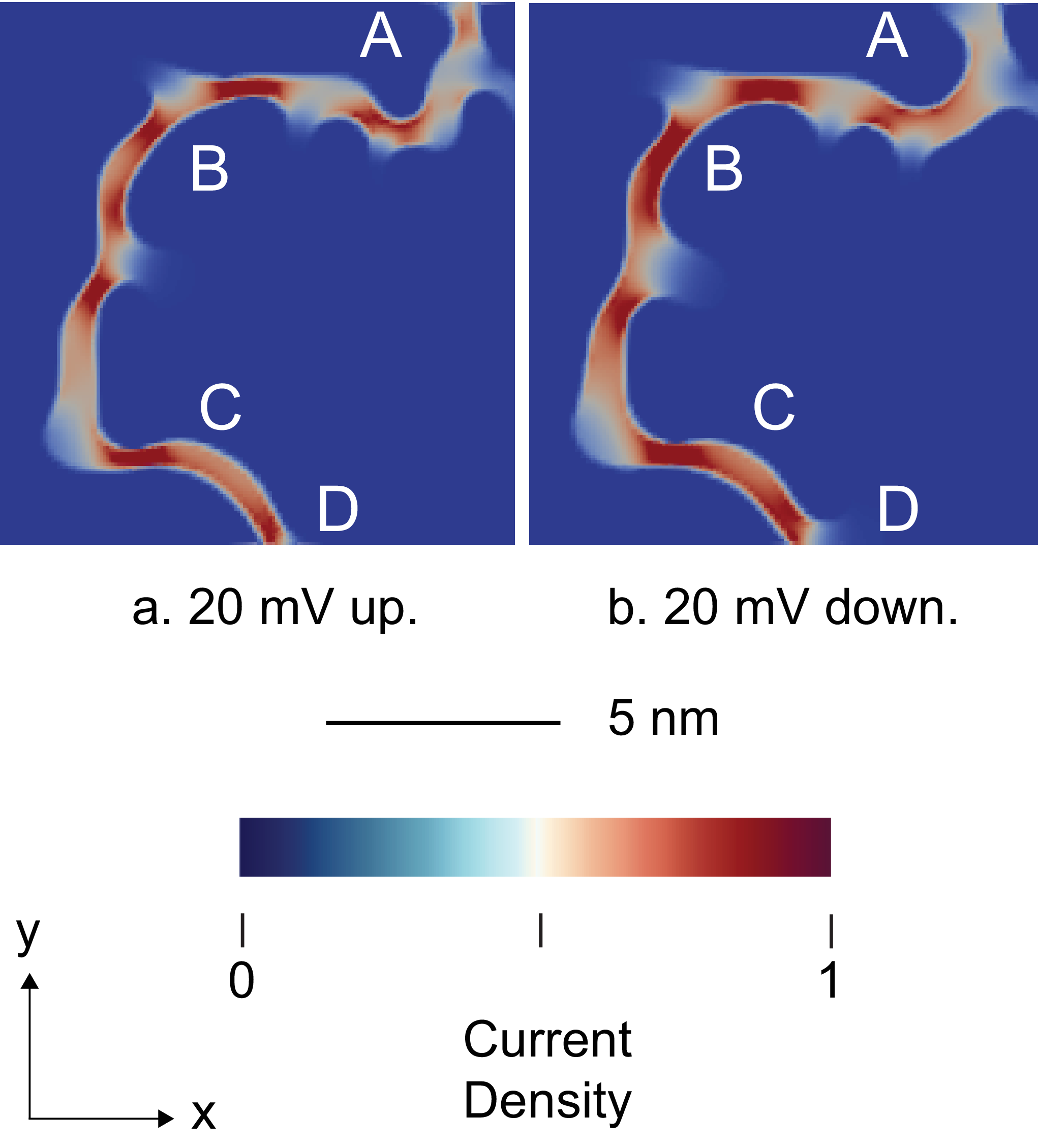}
\caption{Conducting filament morphology and normalized current density magnitude produced by application of bias $V_o$. Panel (a) illustrates steady-state conducting filament morphology for a positive-going bias of 20 mV, corresponding to high resistance state. Panel (b) illustrates steady-state conducting filament morphology for a negative-going bias of 20 mV, corresponding to the low resistance state. See text for explanations of points A-D.}
\label{filaments}
\end{figure}

Figure \ref{filaments} illustrates conducting filament morphology for a positive-going bias of +20 mV and a negative-going bias of +20 mV, corresponding to the high resistance state and the low resistance state of Figure \ref{IV}. The conducting filament morphology follows from two significant features our phase field model. First is conducting filament formation is spontaneous and the morphology does not depend on, nor require, an \textit{a priori} geometry. Second, our method correctly predicts conducting filaments evolve on thermodynamic paths that are energetically favored due to stochastic structural and chemical inhomogeneities naturally occurring at the atomic-level, due to both latent and intentional fabrication effects. Conducting filament morphology of our model is consistent with experimental imaging data \citep{Yang2012}\citep{Ahmed2018}\citep{Bejtka2020}. Clustering is exhibited at points A-D on Figure \ref{filaments}. Comparing point A for each resistance state illustrates substantial a increase in conducting filament diameter as transition to the low resistance state. Similarly, points B, C, and D show a moderate increase in conducting filament diameter as higher energy charges propagate around occupied states.

Our phase field method allows for the animation of dynamic conducting filament evolution. For the web-based version of the manuscript, Figure \ref{filaments} is augmented by a movie that illustrates dynamic conducting filament evolution corresponding to a positive-going logarithmic voltage ramp of 13 steps from 100 $\mu$V to 100 mV and a negative-going logarithmic voltage ramp of 12 steps from 100 mV to 100 $\mu$V along the \textit{current-voltage} response of Figure \ref{IV}.

\section{\label{sec:level2}Summary}

We have presented a multiphysics phase field model inherently compatible with randomness associated with as-fabricated memristive thin films. By treating the motion of oxygen vacancies and their interactions with random states of these amorphous thin films statistically, we avoid the computational complexity of existing methods.

Our method produces \textit{current-voltage} hysteresis with conducting filament morphology consistent with experimental observations. In contrast to previous continuum and phase field methods, no requirements are made on our conducting filament geometry. Our method illustrates conducting filaments spontaneously evolve on thermodynamic paths that are energetically favored due to random structural and chemical variations naturally occurring at the atomic-level due to both latent and intentional fabrication effects. Computational efficiency and scalability of our method is a significant advantage, enabling practical wafer-scale mapping, uniformity, and endurance analysis and optimization.

\bibliography{apssamp}

\end{document}